\documentclass[sigconf,nonacm=true]{acmart}
\renewcommand\footnotetextcopyrightpermission[1]{} 
\usepackage[normalem]{ulem}

\usepackage{smartref}
\usepackage{glossaries}
\usepackage{multirow}
\usepackage{xspace}
\glsdisablehyper

\newacronym{acp}{ACP}{ARM Coherency Port}
\newacronym{asic}{ASIC}{Application-Specific Integrated Circuit}
\newacronym{axi}{AXI}{Advanced eXtensible Interconnect}
\newacronym{bdk}{BDK}{Board Development Kit}
\newacronym{ced}{CED}{Complex Event Detection}
\newacronym{clb}{CLB}{configurable logic block}
\newacronym{crc}{CRC}{cyclic redundancy check}
\newacronym[longplural={Deterministic Finite Automata}]{dfa}{DFA}{Deterministic Finite Automaton}
\newacronym{dma}{DMA}{Direct Memory Access}
\newacronym{fpga}{FPGA}{Field Programmable Gate Array}
\newacronym{fsm}{FSM}{Finite-State Machine}
\newacronym{ewf}{EWF}{ECI Wire Format}
\newacronym{eci}{ACCI}{A Customizable Caching Interface}
\newacronym{ila}{ILA}{Integrated Logic Analyzer}
\newacronym{ist}{IST}{in-system testing}
\newacronym{lut}{LUT}{Look-Up Table}
\newacronym[longplural={Nondeterministic Finite Automata}]{nfa}{NFA}{Nondeterministic Finite Automaton}
\newacronym{regex}{RegEx}{regular expression}
\newacronym{rtl}{RTL}{register transfer language}
\newacronym{rv}{RV}{Runtime Verification}
\newacronym{scu}{SCU}{Snoop Control Unit}
\newacronym[longplural={Systems-on-Chip}]{soc}{SoC}{System-on-Chip}
\newacronym{srga}{SRGA}{self-reconfigurable gate array}
\newacronym{tessla}{TeSSLa}{Temporal Stream-based Specification Language}
\newacronym{vc}{VC}{virtual channel}
\newacronym{io}{IO}{Input/Output}
\newacronym{mpsoc}{MPSoC}{Multiprocessor System-on-chip}
\newacronym{ocla}{OCLA}{On-Chip Logic Analyzer}

\newcommand{\etal}{\textit{et al.\xspace}}
\newcommand{\eci}{\gls{eci}\xspace}
\newcommand{\enzian}{Enzian\xspace}
\newcommand{\cpu}{ThunderX-1\xspace}

\newcommand{\reffig}[1]{Figure~\ref{#1}}

\definecolor{navyblue}{rgb}{0,0,0.502}

\setcopyright{none}

\begin{document}

\title{
ECI: a Customizable Cache Coherency Stack \\ for Hybrid FPGA-CPU
  Architectures
}

\author{Abishek Ramdas}
\affiliation{%
	\institution{Systems Group, D-INFK, ETH Zurich}
	\city{Zurich}
	\country{Switzerland}
}
\author{Michael Giardino}
\affiliation{%
	\institution{Systems Group, D-INFK, ETH Zurich}
	\city{Zurich}
	\country{Switzerland}
}
\author{Runbin Shi}
\affiliation{%
	\institution{Systems Group, D-INFK, ETH Zurich}
	\city{Zurich}
	\country{Switzerland}
}
\author{Adam Turowski}
\affiliation{%
	\institution{Systems Group, D-INFK, ETH Zurich}
	\city{Zurich}
	\country{Switzerland}
}
\author{David Cock}
\affiliation{%
	\institution{Systems Group, D-INFK, ETH Zurich}
	\city{Zurich}
	\country{Switzerland}
}

\author{Gustavo Alonso}
\affiliation{%
	\institution{Systems Group, D-INFK, ETH Zurich}
	\city{Zurich}
	\country{Switzerland}
}
\author{Timothy Roscoe}
\affiliation{%
	\institution{Systems Group, D-INFK, ETH Zurich}
	\city{Zurich}
	\country{Switzerland}
}

\date{}

\begin{abstract}
	
	Unlike other accelerators, FPGAs are capable of supporting cache coherency, thereby turning them into a more powerful architectural option than just a peripheral accelerator. However, most existing deployments of FPGAs are either non-cache coherent or support only an asymmetric design where cache coherency is controlled  from the CPU. Taking advantage of a recently released two socket CPU-FPGA architecture, in this paper we describe \eci, a flexible implementation of cache coherency on the FPGA capable of supporting both symmetric and asymmetric protocols. \gls{eci} is open and customizable, given applications the opportunity to fully interact with the cache coherency protocol, thereby opening up many interesting system design and research opportunities not available in existing designs. Through extensive microbenchmarks we show that \gls{eci} exhibits highly competitive performance and discuss in detail one use-case illustrating the benefits of having an open cache coherency stack on the FPGA.
	
\end{abstract}
\maketitle

\section{Introduction}
\label{sec:introduction}
\glsreset{eci}

The multicore era sparked debate about
whether alternatives to coherence were required as core count
increased \cite{DeNovo11,Cohesion11}, or whether cache coherence could simply
continue to scale \cite{CoherencyScales12}. Its impact on performance
\cite{Hackenberg:Comparing:2009} is as yet unresolved. Yet, with the proliferation of accelerators 
such as FPGAs, the question is as relevant today as it was then, and remains open both in terms of its impact
on functionality and performance as well as in which protocol is most
suitable in which context.

 Based on experience with multicore systems one
might intuitively assume that cache coherence is always a good feature to
have, but on close examination coherence across heterogeneous
components is a difficult and complex topic with many implications for
performance and ease of use \cite{CCPrimer20}.   The ``right'' solution at this point is
far from obvious, and the benefits of coherence in different systems,
workloads, and use-cases are poorly understood and not yet sufficiently explored to make any data-backed claims.

This paper takes a fresh perspective on coherence in a heterogeneous
system.  Rather than asking whether coherence is the right or wrong
approach, we step back and pose a different question: what 
advantages can different \emph{parts} of a representative coherence
protocol \emph{implementation} offer heterogeneous accelerators in a
system? 
We argue the heterogeneity and flexibility provided by
accelerators, especially FPGAs, is too large for a single approach to
coherence to make sense across all use cases, an argument that has also been made for, e.g., GPUs \cite{CCPrimer20}.
In addition, in FPGAs one must be careful with the physical space needed to support cache coherency, which often involves complex protocols with hundreds of states. 

There are several ongoing efforts to standardize connections
between accelerators and CPUs (e.g., Gen-Z \cite{gen_z}, CCIX 
\cite{ccix}, CXL \cite{cxl}).  
These are closed
systems in the sense that they provide a single model of coherency
with few ways to explore options, alternative designs, or opportunities
to tailor the protocol to particular applications.
Moreover, they impose an additional protocol translation between CPU and accelerator, 
rather than talking directly to the CPU's
cache controller. 
Most importantly, however, these standards conflate
separate concerns: cache coherence, caching itself, and communication 
with the accelerator.  
Entangling these aspects necessarily leads to 
compromises in the design, and consequent inefficiencies, not to mention
lost opportunities for more direct interaction. Moreover, they make the protocol more complex, increasing the difficult of efficiently implementing it on an FPGA. A key insight in this paper is that, by separating these three concerns, it is possible to implement an efficient and effective cache coherency stack on the FPGA that allows it to fully interact with the CPU as an equal. 

Based on this insight, we have developed an \emph{open cache coherency
protocol} that uses minimal FPGA resources and shows competitive performance. The resulting artifact 
provides an excellent example of how to customize the cache coherency protocol to particular applications, a feature that will enable researchers and developers to manipulate different aspects of coherency to tailor it to concrete applications. The protocol, \gls{eci}, has been implemented on the recently-released open-source research computer Enzian \cite{Cock:Enzian:2022} which has an FPGA (Xilinx XCVU9P) located in one of the two sockets of a server-class heterogeneous
computer system (ARM ThunderX-1).  \gls{eci} is fully compatible with the CPU's native, inter-socket coherency
implementation and thus provides a realistic and fully-functional cache coherency stack.

In this paper we focus on the design of \gls{eci}, its efficient implementation, and explore a concrete use case that demonstrates the potential of \gls{eci} as a tool to implement coherent heterogeneous CPU-FPGA systems.  The use case we explore focuses on offloading near-memory-processing operations of increasing complexity (sequential access plus filtering, pointer chasing, and regular expression filtering) to the FPGA.
This effectively turns the FPGA into a smart memory management unit, treating FPGA memory as memory in a different NUMA node and returning results directly into the L2 cache of the requesting core, much as a read or write operation over conventional memory would do. This use case offers a nice contrast to existing work implementing similar functionality in restricted settings such as the garbage collection accelerator implemented as part of a RISC-V processor architecture \cite{Maas:2018:Tile-Link-GC} or ACCORDA, a near-memory accelerator prototyped on an FPGA and intended to be inserted on the path between caches and CPUs to offload SQL data processing \cite{Fang:2019:ARDA}. 
For reasons of space, we leave other, more complex use cases that involve manipulating coherency for future work. The use case we analyze in detail demonstrates the feasibility of having a fully compatible implementation of the CPU cache coherency protocol on the FPGA, how tailoring the protocol to the task simplifies it considerably, and the overall performance advantages of \gls{eci} over the bulk data transfer models commonly used today. The idea that a cache coherency protocol can be used for more than just maintaining coherency is starting to gain attention, in particular in relation to accelerators \cite{Calciu:2019:PBerry,CONDA19}. 
\gls{eci} provides an open protocol that boosts such efforts by enabling deeper access to the cache coherency protocol and exploring what cache coherency levels or protocol parts are useful for accelerators.



\section{Background}
\label{sec:related_work}

In this section, we summarize existing CPU
coherence protocols~\cite{Marty:Cache:2008,CCPrimer20}, examine 
how coherent accelerators and
heterogeneous systems affect coherency protocols, and look at the types of
applications often accelerated and how they benefit from different coherency models.

\subsection{The past and present of coherence}
\label{sec:related_work:coherency}


Coherence protocols fall into two classes: \textit{snooping} \cite{Ravishanicar:Cache:1983} and
\textit{directory-based}
\cite{Tang:Cache:1976,Censier:Solution:1978}.  
Snooping protocols are implemented in symmetric multiprocessor (SMP)
systems where all cores share a bus. 
They are broadcast protocols: whenever a core requires a
cache line, it makes a coherence request on the shared bus which is
broadcast to all cores. The remaining cores' coherence controllers 
check if they have a copy of this cache line in their cache and
take appropriate action to maintain coherence. 
Snooping protocols 
suffer from scalability and performance limitations
because broadcasting becomes expensive when more coherence controllers
are added to the system. 

Directory protocols are implemented in coherent NUMA systems.  
Whenever one of the NUMA nodes requires a cache line, it
unicasts a request only to the node that is the ``home'' for that
cache line.  The home node contains a directory that stores
information on  which nodes have a copy of this cache line and in
what states.  The home node then communicates with just the owner 
or sharers to ensure coherence.  Directory protocols do not
require broadcasting messages and so can scale
efficiently but latency can suffer as additional messages need to be
sent via the home node. Systems such as AMD’s Coherent HyperTransport 
and HyperTransport Assist, or Intel's QPI implement versions of directory protocols \cite{CCPrimer20}. 

 Coherence protocols can also be classified by how they handle writes to a local copy of a cache line \cite{CCPrimer20}: \textit{Write-invalidate}
protocols, like MSI, MESI, MOESI and MESIF, ensure a single writer by
invalidating the copies of a cache line in all other caches. New
coherence requests must be made if another core wants to read the
updated copy of the cache line. In a
\textit{write-update} protocol, whenever a core wants to write to a
cache line, it issues a coherence request to update all copies of this
cache line. Although write-update protocols reduce the latency of
reading a newly updated cache line, they are difficult to scale because
update messages are larger than invalidate messages.  As a result
write-invalidate protocols are more common.

The most basic protocol, MSI, consists of three stable states
\textit{modified}, \textit{shared} and \textit{invalid}.  A cache line
in modified state indicates the cache has the only copy of the cache
line in the system and it can therefore be modified. 
A shared state means that the cache has a read-only copy. Other copies may
exist, but must also be in shared state.
The MESI, MOSI, MOESI, and MESIF protocols are optimizations
of MSI considering additional states.  MESI adds an \textit{exclusive} state to indicate that the
cache holds the only copy of the cache line in the system and the copy
is clean.  
MOSI, also called the Berkeley protocol \cite{Katz:1985:MOSI}, adds an additional
\textit{owned} state.  
MOESI, perhaps the most
commonly implemented variant, includes both exclusive and owned
states.  MESIF \cite{Kanter:Common:2007} was developed by Intel and extends the
MESI protocol with a \textit{forward} state.  

In practice, race conditions and concurrency mean that real
implementations have many hidden, intermediate states which greatly
complicate the protocol --- it is not unusual for a coherency protocol
in a multisocket system to have more than 100 states.  As a result,
without exception, these protocols all target CPU-based systems where
multithreaded software shares a single, coherent virtual address
space. If taken as they are, it is likely that an implementation of such protocols on an FPGA would be both inefficient and require to much real state, an argument often made when chosing an asymmetric protocol over a symmetric one (see below) or removing cache coherency completely from the FPGA. 

In addition to their inherent complexity, the traditional properties of standard peripheral interconnects like
PCIe have until recently led accelerators to avoid participating in
these coherence protocols, instead using a computational model like
OpenCL or CUDA based on explicit or implicit copying of the data.
However, even for CPU-only software, the complexity of such protocols
and the difficulty of effectively optimizing performance for them have
led to proposals for, e.g., operating systems that avoid requiring cache
coherence~\cite{Baumann:Barrelfish:2009} and communicate between cores via
messages, which themselves are using the coherence protocol in
unexpected ways~\cite{Giacomoni:FastForward:2008}. 

\subsection{FPGAS and Coherence Protocols}
\label{sec:related_work:accelerators}

A recent trend is for accelerators to be made cache-coherent with the
CPUs in a heterogeneous system.  We focus on FPGA-based
accelerators, but the trend with GPUs is similar \cite{CCPrimer20}. 

Xilinx Zynq Ultrascale+~\cite{Xilinx:Zynq} and Intel Agelix~\cite{Intel:Agilex} are embedded \gls{mpsoc} architectures that combine a
CPU and FPGA on a single die, simplifying the interconnect and offering both
coherent and non-coherent ports between the two. Coherent access is generally
asymmetric, and limited to allowing the FPGA to access the CPU's last level
cache.

Two examples of an \glspl{fpga} connected
coherently with an asymmetric protocol are Intel's HARPv2 system
\cite{Harp_oliver} and IBM's CAPI
\cite{Stuecheli:2015:CAPI}. In HARPv2 a 14 core CPU is connected
coherently to an \gls{fpga} via Intel's Quick Path Interconnect
(QPI). Although QPI supports a symmetric protocol \cite{Intel:QPI},
the \gls{fpga} implements only a caching agent. The coherence protocol
is configurable via a cache coherence table but no information is
available on how to modify the protocol. IBM's CAPI system combines a
12 core processor asymmetrically with an \gls{fpga}. This system is
built on top of PCIe interconnect with coherence being enabled by a
PCIe Host Bridge (PHB) and Coherent Accelerator Processor Proxy (CAPP)
on the processor and a POWER service layer (PSL) on the
\gls{fpga}. Both systems provide a unified addressing space for
applications on the \gls{fpga}.  

The need to coherently connect hetergeneous accelerators to CPUs has led to
several recent interconnect standards.  CXL \cite{cxl} builds coherence and
memory semantics on top of a PCIe transport. It provides a unified coherent
memory space between host CPU and accelerators, using an asymmetric protocol
on the accelerator with a coherence bypass to allow direct access to unshared
parts of device memory. OpenCAPI \cite{Stuecheli:2018:OpenCAPI} also implements an asymmetric
protocol over PCIe (and, more recently, Bluelink). Accelerators are
required to work with caching enabled and virtual addresses, with
virtual to physical mapping done by the CPU's MMU. CCIX \cite{ccix}, on the other hand, supports a symmetric protocol on
the accelerator by extending PCIe. This enables 
accelerators to work as peer of the CPU in an unified, coherent
memory space.  

The above examples favor near-memory processing where accelerator memory 
expands CPU system memory. GenZ~\cite{gen_z} instead
supports both near- and far-memory processing by defining memory semantics for
communication between CPUs, accelerators, and memories. GenZ is designed to
scale to entire racks.
It does not specify operation of the cache coherent agent but
specifies protocols that support building these agents.   

In practice the role of the \gls{fpga} in a coherent CPU-\gls{fpga}
system depends on the protocol implemented in the \gls{fpga}.
\textit{Asymmetric} protocols mean the CPU and \gls{fpga} have a
host-device relationship where the CPU implements both home and
caching agents and \gls{fpga} implements only the caching agent.  To
access pages in \gls{fpga} memory that are marked as 
shared with the CPU, the \gls{fpga} has to make a request to the CPU
which maintains coherence.
In contrast \textit{symmetric} protocols enable CPU and \gls{fpga} to
function as peers by implementing a home and caching agent on both,
exactly as in a homogeneous NUMA system.
The rationale for asymmetric protocols is that the critical access
class for the \gls{fpga} is from device engine to device memory, and
data can be copied in bulk in advance of computing on it.  Symmetric
protocols provide a seamless integration between the CPU and
\gls{fpga} but are naturally more complex to implement.   Both
classes, however, as still ``all or nothing'' cache coherence
protocols. 

\subsection{Accelerated Applications}
\label{sec:related_work:applications}

There has been a massive
proliferation of hardware accelerators with different models of
execution. The initial generations of accelerated applications focused
primarily on the host-device computational model where a host CPU is 
responsible for allocation and scheduling of accelerator resources on external
devices. In this familiar execution model, data is offloaded in bulk, often
via PCIe, for processing and the results are returned to the host.  This
remains the primary operation mode of GPU-accelerated applications (e.g. CUDA~\cite{Nickolls:Scalable:2008} and OpenCL~\cite{Stone:OpenCL:2010}, as well as
more modern application-specific hardware accelerators such as TPUs~\cite{Jouppi:2017:IDPA} and VCUs~\cite{Ranganathan:2021:WSVA}.
This model of execution has not shown much need for cache coherence, not least
because of a historical lack of hardware caches and shared memory.  However,
more generally, the fixed nature of execution provides for efficient
structured offloading, splitting execution evenly between host and device.
Implicit acceptance of this model can be seen in the use of OpenCL to program FPGAs, thereby reinforcing the bulk data loading model for accelerators. 

\section{ECI design} \label{sec:ECI_CC}

\subsection{Systems Overview}

The design of \eci is driven by the emergence of the application classes
above, executing largely non-CPU devices, but which would nevertheless likely
benefit from coherence. That these are \emph{not} CPUs informs our choice to
emphasize flexible protocol subsetting and to expose low-level primitives:
Firstly (in combination with \enzian) to drive the exploration of suitable
coherence models, and secondly as a recognition that specialised hardware may
well need specialised protocols.

\eci is carefully designed to interoperate cleanly with an existing native
coherence protocol to allow realistic evaluation of server-scale workloads,
while remaining scalable both down to specialized sub-protocols and up to new
and more capable platforms.  The implementation is (and will remain) a work in
progress. To demonstrate what performance can be expected from a
carefully-optimized implementation (and to show that it is competitive with
the native protocol), we implement representative prototypes of several
workloads in the currently best-optimized configuration: read-mostly operation
offload to the FPGA, with results in \autoref{sec:experiments}.

\subsection{The Protocol}

The \eci design has several goals: First, to interoperate cleanly with the
native coherence protocol of the \cpu. Second, a clear specification
\emph{independent} of particular machine details.  Third, to be fast. Fourth,
to be extensible to new hardware. Last, it must be modifiable---in particular
to allow protocol subsets for specific applications.

The \cpu implements a 2-node MOESI protocol with home-based directory.
Compared to MESI, dirty lines are forwarded between caches without writing to
RAM thanks to the O (dirty and shared) state.  In a home-based system,
consistency is the responsibility of a single node (the home), which is
responsible for all reads from and writes to the backing store (e.g.~RAM).  If
the remote holds a line in a `clean' state (e.g S), the home node ensures that
the system behaves exactly as though the home node also held a clean copy (S),
or not at all (I). If home holds the line dirty (O), this is completely
invisible the remote node.

Full MOESI is beyond what existing applications are likely to benefit from,
and at the same time a very ambitious target for a first release.  However, we
do not exclude the option of extending \eci to a full MOESI (or MESIF or any
other protocol) in the future.  For now, we opt for a compromise: Beginning
with the native protocol on the ThunderX-1, we abstract the core features
useful for current and near-future applications. We then generalize by
specifying a core set of states and transitions that an implementation
\emph{must} support, and an envelope of things it \emph{may}.

\eci is specified as transitions and messages, agnostic to the underlying
transport. On the \cpu this is a transport protocol guaranteeing reliable
delivery, with multiplexed virtual circuits to guarantee deadlock freedom.
The transport layer will naturally be dictated by the application scenario. It
need only guarantee reliable delivery and a mechanism to avoid deadlock due to
delivery order.

Likewise, the protocol envelope does not specify additional intermediate
states (and associated messages) needed to handle message reordering and
races. As described below, our reference implementation implements all
intermediate states for CPU interoperability, but the user need only consider
the specified \emph{stable} states, even when working with a protocol subset.

\subsection{Protocol Envelope} \label{sec:ECI_CC:envelope}

\begin{figure}
  \centering
  \includegraphics{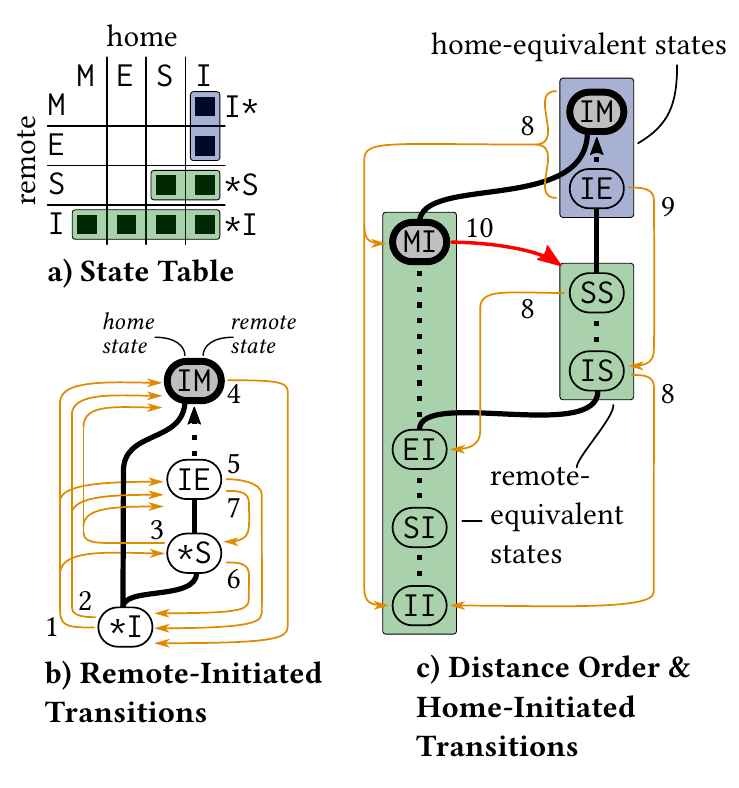}
  \caption{ECI protocol states and transitions\label{fig:eci_states}}
\end{figure}

The home-node protocol of the \cpu gives a natural ordering of protocol states
in terms of the `distance' of data from its at-rest position (e.g.\ DRAM, or
custom query logic for dynamically-generated data), shown in
\autoref{fig:eci_states}.

As shown in \autoref{fig:eci_states}{(a)}, we abstract the
underlying MOESI protocol to an `enhanced' MESI, with the standard set of
allowable state pairs.  The ordering relation between (joint) states is
depicted by thick black lines to states higher in the figure and connected
by either a solid line (for globally-visible transitions) or a dotted line
(for local transitions).  The order is transitive, and thus \texttt{IM}
(invalid at home, modified at remote) compares higher than \texttt{II}
(invalid at both). Two transition classes exist: Upgrades, moving
higher in the distance order (e.g.~transferring from home to remote, or
a line becoming dirty) and downgrades, moving downwards (e.g.~writebacks).

States related only by local (dotted) links are indistinguishable to the other
node, as indicated by shaded rectangles in \autoref{fig:eci_states}{(a,c)},
with the exception that it may be possible to discover \emph{after the fact}
which state the other node was in after a transition e.g.~transition 8
(downgrade remote to invalid) reveals whether the remote node held a
dirty copy (by returning the dirty data) and may thus leave the home node with
a line in either state \texttt{M} or \texttt{I}.

\autoref{fig:eci_states}{(c)} shows all possible combinations of home (left)
and remote (right) states for a cache line e.g.~\texttt{IS} where home node no
copy (\texttt{I}), and remote holds a shared but clean copy (\texttt{S}).  The
thin orange arrows in this subfigure depict the minimal set of transitions
that the home node may generate, and the remote node must therefore support.
These are the \emph{home-initiated} transitions, and only include downgrades:
in the current protocol there is no mechanism to transfer data to a remote
node without that node first requesting it.  This restriction is necessary so
that the baseline specification is satisfied by the \cpu, which lacks any such
mechanism.

\begin{table*}[]
	\begin{tabular}{@{}llllll@{}}
		\toprule
		Initiated by & Transition class & \begin{tabular}[x]{@{}l@{}}Transition\\Request\end{tabular} & \begin{tabular}[x]{@{}l@{}}Request\\has Payload\end{tabular} & \begin{tabular}[x]{@{}l@{}}Response\\from Partner\end{tabular} & \begin{tabular}[x]{@{}l@{}}Response\\has Payload\end{tabular} \\ \midrule
		Remote    & Upgrade         & Read-Shared                      & No                  & Yes                   & Yes                  \\
		Remote    & Upgrade         & Read-Exclusive                   & No                  & Yes                   & Yes                  \\
		Remote    & Upgrade         & Upgrade from Shared to Exclusive & No                  & Yes                   & No                   \\
		Remote    & Downgrade       & Downgrade to Shared              & Yes if dirty        & No                    & No                   \\
		Remote    & Downgrade       & Downgrade to Invalid             & Yes if dirty        & No                    & No                   \\
		Home      & Downgrade       & Downgrade to Shared              & No                  & Yes                   & Yes if dirty         \\
		Home      & Downgrade       & Downgrade to Invalid             & No                  & Yes                   & Yes if dirty         \\ \bottomrule
	\end{tabular}
	\caption{Signalled Transitions.}
	\label{table:signalled_transitions}
\end{table*}

It is permissible for a node to support more transitions than the minimal
subset.  For example, `downgrade remote to invalid and forward' which would
add a transition from state \texttt{IS} to state \texttt{SI} is \emph{not}
included in the minimal protocol (and indeed does not exist on the
\cpu), but would be a potentially useful extension (to avoid a read from
RAM) that an extended implementation might support while remaining consistent
with the basic rules.

Another transition excluded from the minimal (MESI) protocol is that labeled
10 (and coloured red) in \autoref{fig:eci_states}{(c)}, from \texttt{MI} to
\texttt{SI} or \texttt{IS}. This case occurs when a remote node wishes to
upgrade its copy to the shared state (i.e.~to read it), but it's held dirty by
the home node.  This is where the basic MESI protocol is forced to
unnecessarily write the dirty data to RAM rather than simply forwarding
between caches, and is the primary advantage of MOESI.  This transition
\emph{is} allowed by the \cpu, as its native protocol is actually MOESI,
and thus includes the \texttt{O} (owned, dirty and shared) for exactly this
case.  We permit this transition as a concession to performance, with the
requirement that whether the home node's copy is dirty or clean must be
strictly invisible to the remote node. Whether the home node internally uses
the \texttt{O} state or silently writes the dirty data back must not be
visible.

From the ordering on states we derive the following \emph{requirements} for
allowable transitions:
\begin{enumerate}
\item A transition may only occur from a lower to a higher state, or higher to
lower. Transitions between unrelated states e.g. (\texttt{IE} and \texttt{MI})
are forbidden. There is one exception (transition 10), as described above.
\item Any transition between states distinguishable to the other node must be
signaled i.e.~an by exchange of messages. A reply is only required if
necessary for consistency. Transitions over only internal (dotted) edges need
not be signaled.
\item Moving from a dirty to a clean state \emph{must} involve signaling the
home node. This means the edge from \texttt{IE} to \texttt{IM} may only be
traveled upward (as indicated by the arrowhead). The only downgrade path is
thus via the \texttt{MI} path on the home node.
\item States where the remote node holds a clean, shared copy \emph{must} be
indistinguishable to the remote node. If the home node holds dirty data, this
must be entirely invisible to the remote (hidden \texttt{O} state).
\item \label{state_guarantee} Implementations \emph{must not} signal
transitions not supported by the interoperating node.  This rule is in
practice mostly used in the converse: an implementation must support all
transitions the partner may signal, unless it can be guaranteed these won't be
generated (e.g.~with a read-only workload).
\item \label{send_equiv} Any transition a node is permitted to request in a
given state must also be permitted in any other state indistinguishable to
that note (i.e.~reachable by silent transitions of the other node).
\item \label{recv_equiv} A node in a given state must be prepared to receive
any message that it would be prepared to received in any indistinguishable
state.
\end{enumerate}

To ensure reasonable performance, implementations should follow the following
\emph{recommendations}:
\begin{enumerate}
\item Internal transitions \emph{should not} be signalled, particularly
upgrades to dirty states (e.g.~\texttt{IM}).
\item The home node \emph{should} avoid writing dirty lines before sharing
them. This \emph{must} be done in a manner invisible to the remote node.
\end{enumerate}

As already discussed, \autoref{fig:eci_states}{(c)} provides the minimal set
of home-initiated transitions plus MOESI-like remote-initiated
transition 10. Transition 9 is requested by `downgrade remote to
shared', while the three transitions labeled 8 are triggered by
`downgrade remote to invalid'. Where the current state is \texttt{SS} or
\texttt{IS}, the new home state is either \texttt{E} (home now has the only
copy) or \texttt{I}, respectively. In these cases, the home knows which
transition will occur (it can distinguish the two states). The home node
cannot distinguish \texttt{IM} and \texttt{IE} however, as the upgrade to
\texttt{IM} is silent. The final state (either \texttt{IM} or \texttt{II}) is
thus only apparent when the remote node replies, either with dirty data or
without.  For all home-initiated downgrades, the remote node must reply so the
home node can distinguish remote \texttt{I}, \texttt{S} and \texttt{E/M} to
maintain coherency. \autoref{table:signalled_transitions} lists all signaled home and remote transitions.

\autoref{fig:eci_states}{(b)} shows the possible joint states from the
perspective of the remote node, with the remote-equivalent states merged into
the combined states \texttt{*S} and \texttt{*I}. Requirement \ref{send_equiv}
guarantees that this is sound, and with requirement \ref{recv_equiv} that the
remote node need only implement this 4-state protocol. The left-hand side of
the figure shows the mandatory remote-initiated upgrades: Invalid to shared
(1), invalid to exclusive (2), and shared to exclusive (3). The right-hand
side gives the remote-initiated downgrades: Modified to invalid (writeback)
(4), exclusive to invalid (5,6), and exclusive to shared (7).

The MOESI downgrades `modified to shared' and `exclusive to shared' are not
part of the minimal protocol (although the \cpu does permit them).  A
consequence of requirements \ref{send_equiv} and \ref{recv_equiv} is that
remote-initiated (`voluntary') downgrades to either shared or invalid do not
require a reply from the home, as the remote need not distinguish the possible
home states.

\subsection{Specialization} \label{sec:ECI_CC:specialisation}

\begin{figure}[t]
	\centering
	\includegraphics{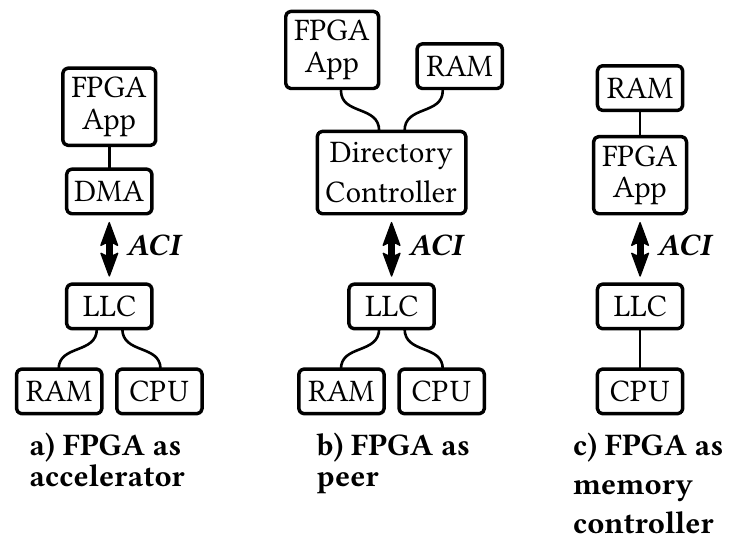}
	\caption{Instances of ECI specialization\label{fig:eci_use_cases}}
\end{figure}

\eci is explicitly intended to be modified to suit particular applications.
\autoref{fig:eci_use_cases} illustrates three examples: (a) an FPGA-implemented
accelerator that interacts with the CPU principally as a DMA initiator;
(b) a fully-coherent symmetric two-node system
(FPGA \& CPU); and (c) a `smart memory
controller' system where most transactions are initiated by the CPU.

Each of these use cases makes different demands on the coherence protocol, and
allow various optimizations and simplifications.  Indeed, for the
CPU-initiator case and a read-only workload, the number of states needed to
be distinguished on the host node (the FPGA for workloads such as operator
pushdown) is one, and the FPGA accelerator can remain coherent despite
implementing neither cache nor directory!  We demonstrate this
optimization is safe, and evaluate its performance in
\autoref{sec:experiments}.

Refer again to \autoref{fig:eci_states}{(b)}, and consider a read-only
workload. For the remote node (here the CPU), the \texttt{IM}
and \texttt{IE} states do not occur, leaving only \texttt{*S} and \texttt{*I}.
Only transitions 1 (upgrade to shared) and 6 (voluntary downgrade to invalid)
remain.  Appealing to requirement \ref{state_guarantee}, the home node
(FPGA) need only respond to the upgrade (which requires a reply) and the
downgrade (which doesn't).

Turning to \autoref{fig:eci_states}{(c)}, we can immediately discard states
\texttt{MI}, \texttt{IM}, and \texttt{IE} and the associated transitions (9,
10, and one case for 8). If the FPGA does not locally cache the data (it will
be cached on the CPU, the FPGA will never need to deal with a dirty line),
then the \texttt{EI}, \texttt{SI}, and \texttt{SS} states likewise vanish,
leaving only a two-state protocol consisting of \texttt{IS} and \texttt{II}
and the host-initiated downgrade to invalid transition (8). The only reason
for this one remaining home-visible transition is to evict data known to be
clean, and it too may be discarded, leaving only the combined state
\texttt{I*} (uncached at home, remote state shared or invalid) and no
host-initiated transitions. The FPGA-side home node need only respond to
`upgrade to shared' requests with the necessary data, and silently ignore
voluntary downgrades from the CPU: neither requires transitioning from
\texttt{I*}, and thus the FPGA need track no state at all for a cache line.

Note that this dramatic simplification does not impact the CPU's ability to
cache the data (often generated at great cost in time and energy by an
accelerator). As \autoref{sec:experiments} shows, not only does the
simplified endpoint interoperate flawlessly with the \cpu, the performance of
workloads with significant temporal locality increases dramatically thanks to
transparent use of the CPU's L1 and L2 caches.

\section{The ECI Toolkit} 
\label{sec:development}


\eci is based upon the existing \cpu coherence protocol. 
Despite excellent support from the CPU
vendor in the form of documentation and conversations with the
designers, coherence on the \cpu was never designed (or
documented) with a heterogeneous system use-case in mind -- the design
goal was for two identical CPUs to talk to each other in a 2-socket
homogeneous system.  We used a combination of tracing technologies,
ad-hoc trace analysis tools, formal specifications of different
protocol layers, and simulation techniques to help both in
implementing \eci and also as a subsequent aid for developers using
it. These tools are availiable to make it easier to explore \gls{eci} in detail. 

\subsection{Supporting tools}

\textbf{Trace capture:}
We started by gathering a set of reference traces from two \cpu
systems configured in a 2-socket NUMA configuration booting Linux from
power-on up to the shell prompt.  By interposing an FPGA board between
the two cores, and reducing the number of 10 Gb/s lanes used by the
coherence protocol, we were able to capture bidirectional traces at
the block level and download them to a PC for analysis.

From these low-level traces we constructed the sequence of coherence messages
in each direction on all virtual circuits used by the protocol.  We
defined our own JSON-based serialization format for these messages along with
a canonical binary format, \gls{ewf}, to allow the decoded traces
to be used for a variety of purposes.  We also wrote a plugin for the popular
Wireshark protocol analyis tool~\cite{Wireshark} for visualizing the protocol.

\textbf{Formal specification and modelling:}
Our evolving view of \eci was captured in a set of formal
specifications which we checked against our traces.  One of these used
the SAIL specification language~\cite{sail} for specifying the format
of the individual messages --- indeed, the code for decoding these
messages into our serialization format was generated from this
specification.

A second specification captured the valid protocol message exchanges.
A cache coherence protocol often has many outstanding transactions.
We used this spec against our traces to check that our state machines
were consistent with the observed behavior of the native protocol ---
when they disagreed, we reworked our specification.

It should be noted that the protocol itself does not simply consist
of coherence-related messages. Non-cacheable I/O accesses, memory
barriers, and interprocessor-interrupts are all carried via this protocol. 

A third specification attempted to capture the actual coherence
protocol itself: state transitions for cache lines, intermediate
states when messages were in flight, etc.  Here we were helped by
considerable prior work on specifying cache coherence protocols in
academia, and the fairly clean nature of the protocol we were
modeling.  The resulting specification was a considerable superset of
that required for \eci, and covered 4-node NUMA systems.

\textbf{Simulation:}
Given ``legitimate'' traces of messages and a standard format for
representing these in software, we could now simulate either end of an
\eci link.  For the FPGA side, we used a standard Verilog simulator in
place of the FPGA to send and receive \eci messages using our evolving
\eci implementation.

For the CPU side, we implemented a custom cache module in C++ for the
ARM Fast Model simulation suite~\cite{fastmodels} which modelled the
\cpu L2 cache.  This allowed us to run application binaries over Linux
as if on the \enzian CPU, but our cache module would also send and
receive \eci messages correctly (modulo our specification) over a
network socket using our JSON format and adjust the cache state
accordingly.

Connecting these two over TCP sockets provided a viable simulation
enviroment for the entire machine, useful not only for debugging \eci
itself, but also for those developing both software and FPGA
applications.

\textbf{Online tracing:}
One deficiency of using offline reference traces to guide the
implementation is that rare, unusual events are sometimes missing from
the traces.  Failing to react on the FPGA side to these events often
caused a machine check on the CPU, with very little information
available as to why.

In addition to the \gls{ocla} on the FPGA, we developed a tool
for online tracing of \eci which could check parts of our protocol
specification against a running \enzian system at the full link rate
of 240 Gb/s and record violations of our
specification.

The parts of the protocol to be verified are specified as
\glspl{nfa} using a simple language, which is compiled into a circuit
synthesized on the FPGA along with the rest of the FPGA configuration.  
Loading a new configuration takes only a few seconds, whereas
resynthesising a specialized engine can take many hours.  The tool
requires modest \gls{fpga} resources, yet can capture complex events
at full line rate without any additional latency.   It can thus be
used to debug and analyze the behavior of very high-speed
interconnects without interrupting execution, enabling the observation
of transient, complex events in real time, and has proved invaluable
in ironing out the remaining interoperability corner cases in \eci, as
well as debugging applications using it.

\begin{table}[t]
\caption{\eci hardware resource consumption, percentage over the resources available in a Xilinx VU9P}
\begin{tabular}{|l|l|l|l|}
\hline
                          & LUTs   & REGs   & BRAM(36Kb) \\ \hline
\eci per link  & 46186  & 32777  & 112.5      \\
Percentage & 3.91\% & 1.39\% & 5.23\%     \\ \hline
\end{tabular}
\label{tab:eciresource}
\end{table}

\subsection{Reference FPGA Implementation}
\label{sec:development:reference_fpga_implementation}

The reference implementation of \eci is inter-operable with ThunderX-1 CPUs,
open, and reconfigurable. The implementation is layered: \gls{vc} layer, link
layer, transaction layer and physical layer. The \gls{vc} layer implements 14
different virtual channels that expose \gls{io} and coherence operations to
the \gls{fpga}, of which 10 are for coherence traffic, with separate sets of
\gls{vc}s for odd and even cache lines enabling simpler load-balancing. In
order to provide a deadlock free network, the current implementation of the
\gls{vc} layer is based on the message classes defined by ThunderX-1 and can
be reconfigured as necessary.

There are no ordering guarantees across \gls{vc}s and the implementation thus
handles race conditions using additional intermediate states, invisible to the
application. The link layer formats coherence messages and efficiently packs
them for transport through lower layers. The transaction layer manages link
state, credit based flow control, and error and replay mechanisms to ensure
delivery of messages. The physical layer is responsible for transport of \eci
messages through serial lanes. All these layers have been tested and have good
performance. In addition, \gls{eci} needs very few resources on  the FPGA, leaving plenty of room for the application (Table \ref{tab:eciresource})

The \gls{fpga} can choose to implement asymmetric or symmetric versions of the
coherence protocol and reference designs are available for both. 
To support the 2-node symmetric model where the CPU and \gls{fpga} operate as
peers with respect to the coherence protocol, an a directory controller implementation
is available which implements a state space
that can be tailored to needs of different applications. For example, the
reference implementation of the protocol is tailored towards expanding the
CPU's memory by being able to access the \gls{fpga}'s memory coherently.
Though not present in the reference implementation, the FPGA can choose to
implement a cache or interact directly with the directory controller to access
both CPU and \gls{fpga}'s memory. The directory-controller's entire state
machine, including intermediate states to handle race conditions, is generated
automatically from a formal specification. For reasons of space, we do not discuss 
the directory controller implementation any further in this paper but the code is available. 

In this paper we explore in more detail the 2-node asymmetric configuration, turning the
 \gls{fpga} into a custom memory controller implementing near-memory-processing operators. 
In this use case, the \gls{fpga} interacts directly with the coherence protocol to function as a custom
memory controller providing different coherent views of the device memory
to the host and placing the data directly on the L2 cache of the requesting core. 
We use this use case to illustrate the performance that can be expected from a fully-optimised
implementation of \eci on an FPGA and its versatility in practice. In this design, the FPGA reads its 
memory on behalf of the CPU but it implements high level operators that process the data in flight between the FPGA memory and the CPU cache. We have implemented three operators corresponding to
common FPGA-offload scenarios in the literature: an SQL \texttt{SELECT} operator
pushdown where the FPGA filters data according to a simple predicate, a pointer chasing operator that can traverse data structures to find the relevant data items, and a regular expression matching engine that is used to filter data before sending it to the CPU.  These are all instances of topology c of \autoref{fig:eci_use_cases}, this being the most
mature of our reference implementations, already achieving close to the
theoretical performance limit.  All three workloads take advantage of the
optimizations permitted by the specialization described in
\autoref{sec:ECI_CC:specialisation}.



\section{Evaluation}
\label{sec:experiments}

We show, first, that \eci's flexibility has essentially no cost, and
that we can implement full coherency on the FPGA.  Next,
microbenchmarks show performance as close to the CPU's native
implementation as is likely possible using an FPGA.  Finally,
macrobenchmarks implementing operators in the memory controller
illustrate the design and provide examples of how \gls{eci} can be
employed in practice.

\subsection{Hardware Platform}

We use the recently-released \enzian research computer
\cite{Cock:Enzian:2022} as our experimental platform: 
\begin{itemize}
\item CPU: Marvell ThunderX-1, 48x dual-issue ARMv8, 2.0GHz, 16MB
  16-way associative LLC, 128B lines.
\item CPU DRAM: 4x 32GiB 2133MT/s DDR4 channels (only 2 used)
\item FPGA: Xilinx Ultrascaloe+ XCVU9P at 300MHz.
\item FPGA DRAM: 4x 16GiB 2400MT/s DDR4 channels (only 2 used)
\item Interconnect: 30GiB/s bidirectional (theoretical, including overheads).
\end{itemize}

\subsection{Microbenchmarks}\label{sec:eval:micro}

We first look at the performance of \eci as an inter-socket
interconnect.  The key question is whether coherency protocols
implemented on FPGAs can realistically interoperate with native CPU
implementations.

We compare throughput and latency of inter-socket cache operations on
\enzian with an off-the-shelf conventional 2-socket server machine
with the same CPUs, which provides an upper bound on the achievable
performance of \eci. 

\begin{table}[t]
  \centering
\begin{small}
  \begin{tabular}{|r|c|c|}
    \hline 
    & \enzian + \eci & 2-socket server (native) \\
    \hline
    Throughput: & 12.8 GiB/s & 19 GiB/s \\
    Latency:    & 320 ns     & 150 ns   \\
    \hline
    \end{tabular}
\end{small}
  \caption{\eci performance comparison}
  \label{tab:microbenchmarks}
\end{table}

The result in \autoref{tab:microbenchmarks} shows that, while our
current \eci implementation could undoubtedly be further optimized,
these results do show that \eci exhibits realistic performance for
cache coherence hardware.  Note that the FPGA has a much lower 
clock speed than the CPU and, as a result, it will always be intrinsically
slower than a CPU executig the same protocol. Nevertheless, we believe that
further optimizations are possible that will reduce the gap between the
CPU and the FPGA implementation.

\subsection{Experimental Setup}
\label{sec:eval:setup}

\subsubsection{Operator interface via \eci}

\begin{figure}[t]
	\includegraphics{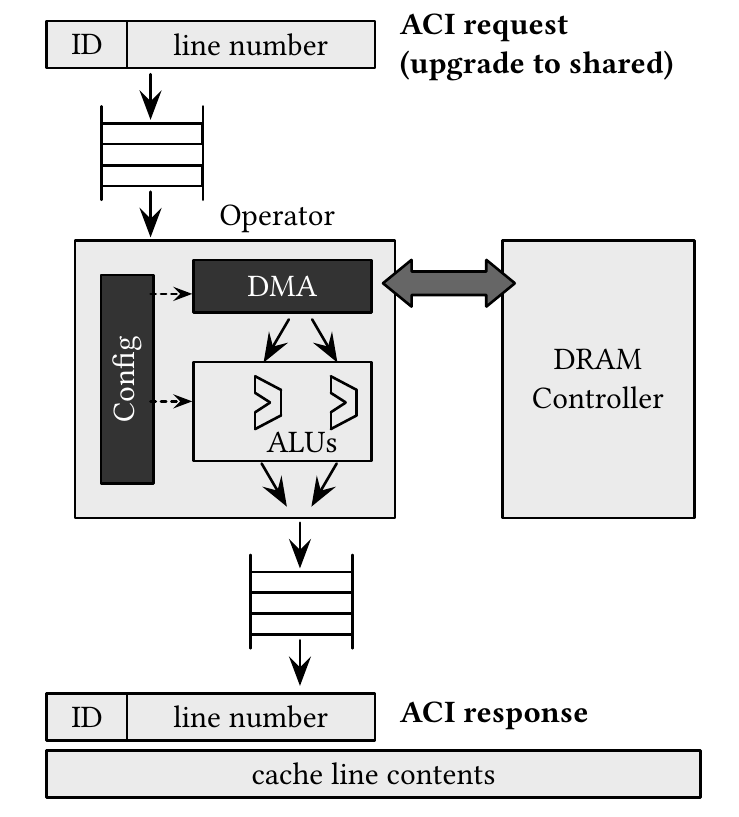}
	\caption{The common structure of experimental operators}
	\label{fig:opmodel_interface}
\end{figure}


The FPGA portion of all three workloads is implemented according to the
structure depicted in \reffig{fig:opmodel_interface}. The operator receives
commands as requests over \eci: specifically, requests to upgrade a cache line
to the shared state. For the select pushdown and regex operators
the line number (i.e. read address) is unused, as both return results FIFO. 
For the pointer-chasing workload the address is hashed to
identify a bucket in which to search for a key.

The main data flow through the operator is from FPGA-side DRAM via an
integrated DMA engine, through the arithmetic units, and out to the CPU's
LLC (L2 for the \cpu) as a response to an upgrade request. The
request flow is from the CPU (as upgrade requests), through the DMA engine and
arithmetic units, and back as a \eci response.  From the CPU perspective, 
the performance-critical portion of the workload is read-only, with the
FPGA acting as a `smart' memory controller, performing data reduction and/or
computation as appropriate. Data is carefully packed to make efficient use of
cache-line-sized transfers (128B).

Each operator is configured by read/write access (also over \eci) to a
\emph{config} module, e.g.\ to set query parameters or to load a regex. 
This communication is not on the critical path of
the workload, and is not included in our measured results.

The CPU cores are mostly in-order, and thus tend to serialize very quickly in
the presence of significant latency, which grows as the FPGA does more work.  
Thus performance scaling is mostly achieved (as the \cpu designers
intended) through parallelization across the many cores.  Both the
\eci implementation and our operators are highly concurrent, and service a
large number of results in parallel using multiple arithmetic units.  The
principal scaling parameter in our experiments is thus CPU thread count.

%
%

\subsubsection{Hardware interface for multi-operator design}

\begin{figure}[!t]
	\includegraphics{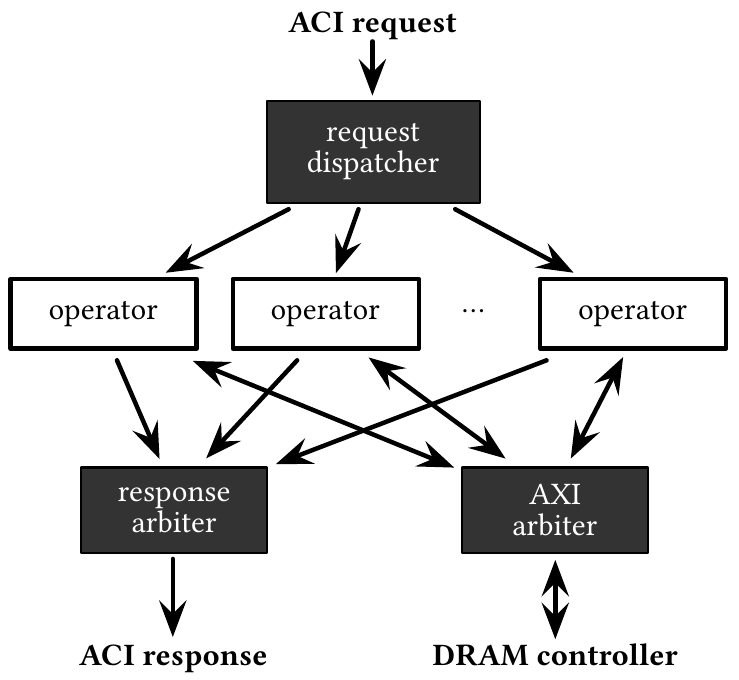}
	\caption{Parallel operator pipeline}
	\label{fig:opmodel_multiop}
\end{figure}

Some workloads, including our pointer-chasing example, are heavily constrained
by the latency of outstanding DRAM requests, which take \texttildelow100ns on
\enzian. The 512b interface provided by the DRAM controllers limits
such an operator to \texttildelow640MB/s throughput.

To achieve higher performance, it is necessary to run multiple
parallel operators, as shown in \reffig{fig:opmodel_multiop}. Here, \eci
requests are fanned out by a central dispatcher to many operators,
each incorporating a DRAM controller. This structure is used in
the pointer-chasing workload.

\subsection{\texttt{SELECT} pushdown}
\label{sec:eval:sel}

The first operator we explore in the FPGA based memory controller is an SQL's
\texttt{SELECT} operator that filters data according to a simple predicate. 
It demonstrates the overall performance improvement
obtainable for data-reduction operators with little or no compute by limiting
the amount of data moved across the machine. It also shows the advantages of 
\gls{eci} for building applications. Because the filtering is transparent, the 
CPU only issues a read operation to a particular address. This mechanism reflects 
the \textit{volcano execution model} used in relational engines where tuples are passed 
from an access method to the rest of the query plan through a series of \textit{next} calls. 
With \gls{eci}, the access method implementing the selection is pushed down 
to the memory controller on the FPGA with matching tuples passed on to the CPU LLC. 

The operator supports queries of the form \texttt{SELECT * FROM S WHERE S.a >
X AND S.b < Y}, \texttt{a} and \texttt{b} are two attribute values of row
\texttt{S}, which is sized to fit within a cache line.  The operator performs
a table scan when triggered by a read from the CPU to a FIFO address, and
returns matching rows in order upon receiving further reads.  Multiple cores
may safely read the FIFO concurrently once the scan is initiated, and will
receive interleaved results.  Matched rows are pushed to an output FIFO and
returned on a first-come first-served basis.  The operator is fully pipelined.
The table contains $5120000$ rows (655~MB). We vary selectivity, the
proportion of returned records, to show the effect of bottlenecks in different
components.

%

\begin{figure}[t]
    \includegraphics{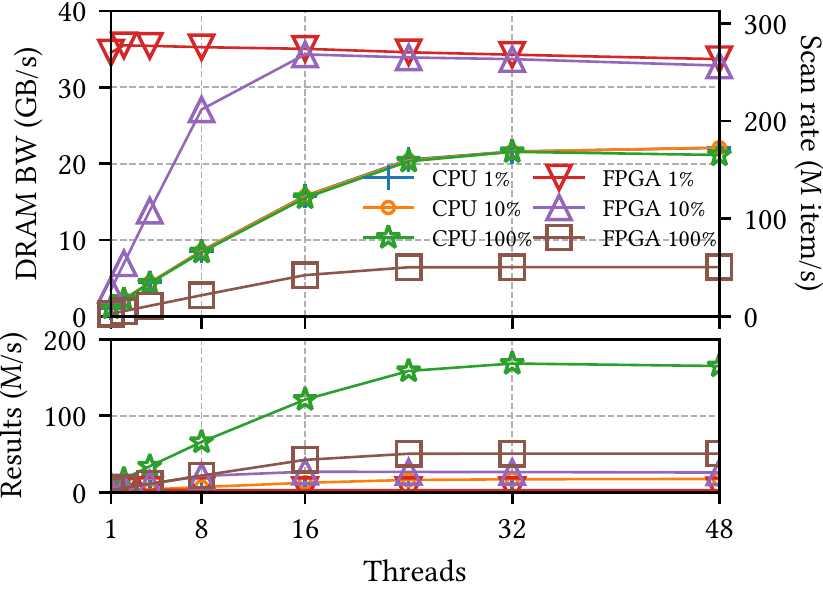}
    \caption{\texttt{SELECT} throughput vs.\ selectivity and thread count for
    CPU and FPGA implementations.\label{fig:sel_comparison}}
\end{figure}

\reffig{fig:sel_comparison} presents the throughput for selectivity
1\%, 10\%, and 100\% against thread count.  The measured quantities are scan
rate (rows/second) and DRAM read bandwidth (which is proportional) in the
top plot, and results returned per second in the lower. The CPU curves are for
an operator running entirely on the CPU with data in CPU DRAM, while the FPGA
curves are for an an operator initiated by the CPU but executed on the FPGA
with data in the FPGA DRAM.

The scan rate on the CPU is independent of selectivity, and is limited by the
CPU's DRAM bandwidth. The FPGA's scan throughput depends strongly on selectivity:
Where the fraction of matching results is less than the ratio of interconnect
bandwidth to FPGA DRAM bandwidth ($1:6$ here), the FPGA
operator is limited only by DRAM bandwidth once enough threads are running to
keep the pipeline full (16 for 10\% selectivity). Once selectivity is high
enough to saturate the interconnect (e.g. at 100\%) the scan behaves as
the CPU implementation, albeit with the reduced bandwidth to remote DRAM.

The curves for results per second show an inversion for high-selectivity
queries, as we would anticipate from the scan rate. When most of the data is
returned to the CPU, higher bandwidth to local DRAM becomes dominant. For
lower selectivities, i.e.,~where there is an effective data reduction, the
FPGA-offloaded operator achieves higher performance, and at a lower CPU thread cost.

\subsection{Pointer chasing in a key-value store}
\label{sec:eval:kv}

The second operator we implement in the memory controller tests the utility of offloading a latency-bounded
operation: pointer chasing, to the FPGA. The workload is a key-value store (KVS)
implemented as a hash table with separate chaining. A key (encoded in the
address sent over \eci) is hashed to select a bucket, which contains the head
pointer to a linked list of key-value pairs. Each (read) request from the CPU
triggers a pointer chase along the linked list to locate the matching key.
Each CPU core will block waiting for a result, and parallelism is achieved
through multiple outstanding requests, using the multiple-operator
architecture described in \reffig{fig:opmodel_multiop}.

Each list entry is 128B, comprising an 8B key, 112B value, and 8B pointer to
the next entry. The KVS contains 5120000 key-value pairs, uniformly
distributed between buckets. To simulate different table fill states we vary
the chain length and search for the last key in the list to force a
known-length pointer chain.  The FPGA implements 32 parallel operators.

\begin{figure}[t]
	\includegraphics{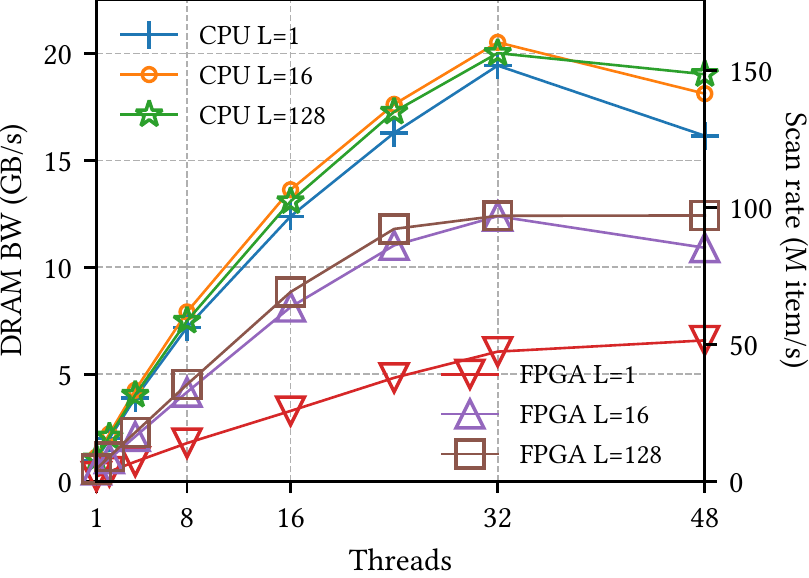}
	\caption{Pointer-chasing throughput on CPU and FPGA for varying chain
    lengths.\label{fig:kv_process_rate}}
\end{figure}

\reffig{fig:kv_process_rate} shows the throughput of the pointer
chase operator, with DRAM bandwidth on the left and keys/second on the right.
The chain length varies from 1 to 128. Clearly, this is a
negative result for this particular workload, but a success for \eci as a
prototyping system: We could quickly determine whether a
seemingly-promising accelerator would perform well in practice. 
The limiting factor here is the random-access performance of
the DRAM subsystem for both CPU and FPGA, and here the CPU has an advantage
with its large cache, higher clock frequencies, and carefully-tuned design. The
length-1 curve for the FPGA shows again the effect of interconnect saturation,
as the number of DRAM accesses per \eci transaction is 1.

\subsection{Regular expression matching}
\label{sec:eval:regex}

The final operator we consider integrates an open-source regular expression matching
engine~\cite{Sidler:2017:Accelerating,Istvan:2016:Runtime,Sidler:RegexRTL:2017} into the memory controller.  
This extends the \texttt{SELECT} operator of
the first experiment to support more sophisticated filters, namely SQL's
\texttt{REGEXP LIKE}, which implements text search over strings using regular expressions.
While still filtering data, the regex
match is dramatically more computationally intensive, which consequently
changes the workload characteristics. This workload thus stands in for the
general offloading of \emph{compute-intensive} operations.

The operator works on a 62B string field within the 128B
row, and takes one cycle-per-character, fully pipelined. Mismatches terminate
early. As it was the case for the \texttt{SELECT} experiment, the operator performs
table scans on behalf of the CPU, with matching rows returned in a FIFO from
which cores read concurrently. The table again contains 5120000 rows, and the
FPGA uses 48 parallel matching engines. We seed the table with a set
number of matching strings to control the query selectivity. The CPU-only
implementation uses a well-optimized open-source library~\cite{Kokke:TinyRegex:2021}. 

\begin{figure}[t]
    \includegraphics{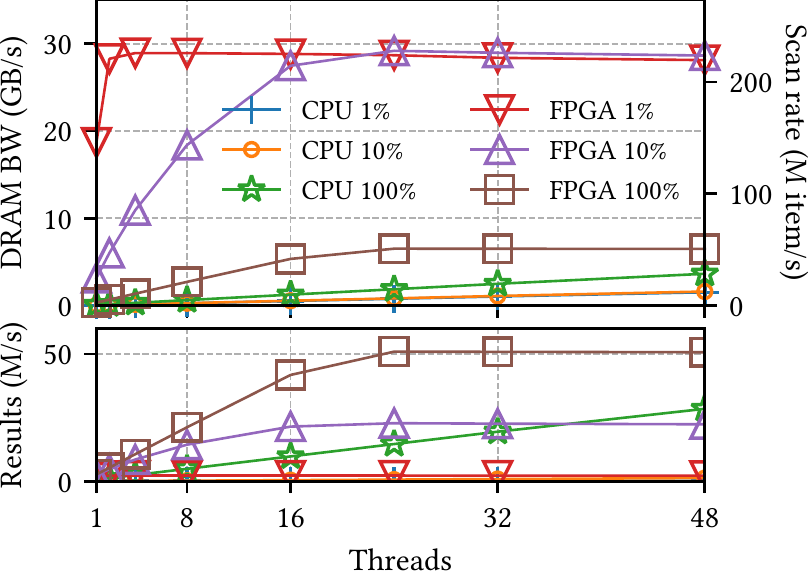}
	\caption{Regular expression throughput vs.\ thread count and selectivity
    for CPU and FPGA.}
	\label{fig:regex_comparison}
\end{figure}

\autoref{fig:regex_comparison} shows the throughput results, reported 
as for previous experiments. The overall pattern is essentially the same,
except that thanks to high computational intensity, and that regex matching 
is particularly well-suited to FPGA implementation, the
FPGA-offloaded version outperforms the CPU in every
case, even when there is an interconnect bandwidth bottleneck in the 100\% selectivity
case. The FPGA implementation achieves twice the performance of the CPU
version (operating on local DRAM) even when every result matches and must, thus,
be sent to the CPU. This excellent throughput is achieved with only a third of the
CPU cores involved (and it is mostly stalled on memory cycles).

\subsection{Exploiting temporal locality}
\label{sec:eval:temploc}

The results of all the operators are delivered to the
CPU's L2 cache in a manner completely invisible to both the CPU software and 
the operator itself, thanks to \eci integration. If a result
delivered to the CPU is expensive to compute in time or energy,
reusing rather than recomputing it represents a significant saving.

To demonstrate the ease with which this can be achieved, with dramatic
performance improvements, we took the baseline regex-matching
scan of the previous experiment, and simulated an enclosing application with
varying degrees of temporal locality. Instead of iterating through the scan
once, we arrange that when the CPU reads value $N$, it will also
re-read value $N-D$, $N-2D$, and so on, up to the size of either the L2 or L1
cache, as appropriate. We can thus directly control the degree of result reuse, 
which in this example are computed at great cost.

\begin{figure}[t]
	\includegraphics{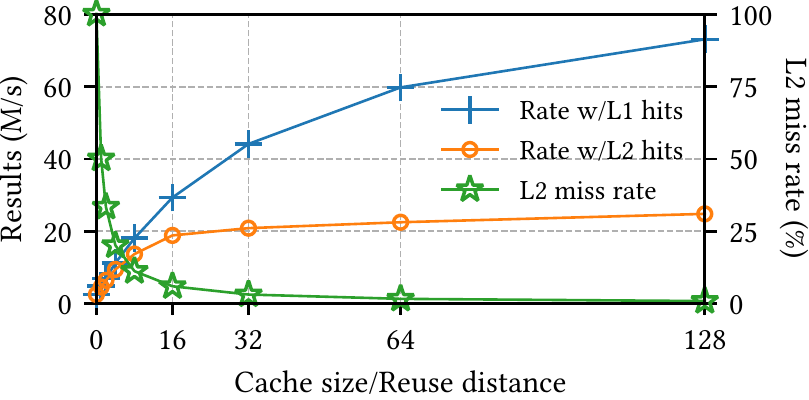}
	\caption{The effect of temporal locality with \eci}
	\label{fig:regex_loc}
\end{figure}

\autoref{fig:regex_loc} shows the result, with controlled parameter $D$
as a fraction of the cache size, approximately equal to the number of
times a result will be reused (modulo boundary effects and cache pollution).
We report two different series: one each spanning the L1 and L2 cache sizes. 
For the L2 case, we plot the miss rate reported by CPU performance
counters. All data was collected with one thread and a selectivity of 10\%.

As can be easily seen, the effect is dramatic, with a single core
outperforming the entire system at a reuse rate of 16 in the L2 cache, and 8
in the L2. This experiment illustrates the potential of
fine-grained coherence for CPU-FPGA applications, and the benefits that a
flexible low-level interface such as \gls{eci} can deliver.

\section{Related Work}
\label{sec:Related}

Cache coherency is well studied in architecture and systems; here we focus on work targeting access and customization of the protocol for accelerators in different settings. 

Berg showed (for coherence between CPU and device)
the benefits of different coherence modes (software, hardware, application-controlled scratchpad) vary significantly with application
and system characteristics and thus no single protocol is necessarily ``correct" \cite{Berg:Maintaining:2009}.
This observation, albeit based on traditional I/O-based devices,
agrees with our contention that in non-CPU-like devices,
the ideal coherence protocol depends on the application, with
accelerators, the execution unit. 

For FPGA SoCs, cache coherency has been studied using Networks-on-Chip
(NoCs)~\cite{Giri:Accelerators:2018,GiriNOCS}. 
For example, the Zynq MPSoC provides the AXI-based \gls{acp} interface
which allows for coherent accesses using the ARM \gls{scu}.
Sadri \etal~use this to evaluate power/performance tradeoffs in
various configurations of an image processing
application~\cite{sadri2013energy}. 
 Similarly, TileLink
\cite{Cook:CARRV:2027} is a chip-scale interconnect protocol for
RISC-V processors providing coherent access to shared memory, using a
customizable, MOESI-equivalent protocol optimized for tightly-coupled,
low-latency SoC buses.
These single-chip platforms target rather different use cases to the
heterogeneous CPU-FPGA platforms we address.  While they also open up
the protocol, \gls{eci} is an actual protocol stack fully
compatible with the CPU coherence protocol, targeting server-class
systems.

HARP was a server-class platform allowing coherent
applications~\cite{Harp_oliver}.  Unlike \gls{eci}, HARP used 
QPI and only implemented an asymmetric protocol with limited access to
the FPGA.  Cabrera and Chamberlain ported OpenCL kernels from
PCIe-based platforms to HARP, showing the benefits of shared
virtual memory over explicit reads and
writes~\cite{Cabrera:2019:EPPO}. The Centaur shell for HARP is a
framework running hybrid relational queries, dynamically allocating
FPGA operators to query plans~\cite{Owaida:2017:Centaur}. 
Moreover, machine learning
applications~\cite{Owaida:2019:DIDT,arnold2020single} demonstrated
coherence-related speedups on HARP over CPU-only implementations.
These show the advantages of FPGA participation in the coherency
protocol, which can be enhanced by the FPGA actively participating in
a symmetric protocol, as with \gls{eci}.   

CAPI (see \autoref{sec:related_work:accelerators}), is another asymmetric 
protocol allowing the accelerator to cache data.
Ito and Ohara build a bioinformatics algorithm (pair-HMM) on a
CAPI-enabled system, achieving 33x better power-performance than with
a POWER8 CPU alone~\cite{ito2016power}.  They conclude that coherence
between CPU and accelerator makes programming simpler and more
efficient than with traditional accelerator design, a finding echoed
by Mughrabi \etal accelerating PageRank algorithms by leveraging
coherent caching on the FPGA~\cite{mughrabi2021qpr}. 
Van Luntern \etal~ accelerate stencil processing using both
a FPGA and ASIC in tandem with POWER9~\cite{van2019coherently} via
OpenCAPI. These all show significant power and performance benefits
of various coherence models, while enabling programmers to use
both existing and novel caching and threading techniques, and
motivated our development of \gls{eci}. 

The use case we explore in this paper is similar to recent work on
near-memory processing.  Maas et al.\cite{Maas:2018:Tile-Link-GC} 
implement an on-die garbage collection accelerator for RISC-V processors
connected to shared memory through TileLink. Similar functionality can
be developed on \gls{eci} without changing the processor
architecture.  This shows a fundamental difference 
between \gls{eci} and protocols like TileLink:
\gls{eci} enables a reconfigurable accelerator to interact with the CPU
at the coherence level, but crucially can be reconfigured on a
per-application basis, whereas SoC or processor designs cannot be
viewed as a dynamic component of the application: \gls{eci} has been
developed to support software applications rather than processor
components.

Similarly, the ACCORDA~\cite{Fang:2019:ARDA} proposal for integrating
a specialized SQL operator engine into the memory hierarchy, and the
Oracle SPARC M7 ``Software in Silicon'' accelerator, operate on the
stream of data between DRAM and processor cache, while \gls{eci}
connects a programmable accelerator for arbitrary tasks to the
coherent interconnect.  Nevertheless, there are similarities which
show the value of the functionality all these systems provide.

\gls{eci} is the first design to provide a coherence
stack fully compatible with a server-class CPU in a two-socket
architecture, with few restrictions on the accelerator functionality -
indeed, \gls{eci} enables applications dynamically deployed on the FPGA
tailor the protocol at runtime, transparently to the CPU.

\section{Conclusions}
\label{sec:conclusions}

\eci is 
an
open protocol implementation running on accelerators (currently FPGAs)
and inter-operating with the CPU's native cache coherence
implementation. \eci presents an abstracted view of the protocol to
accelerator applications which allows subsetting the full protocol
to achieve better performance and space efficiency, while insulating
them from the full complexity of inter-operating with the CPU.
 \eci has performance comparable to the original
inter-CPU implementation with which it inter-operates, and we illustrate potential 
applications that benefit from \eci.

Future work includes further \eci implementation optimizations and exploring a number
of intriguing scenarios enabled by \eci: streaming data arriving from 
the network directly to the CPU cache, extending coherency across networks of FPGAs 
connected via RDMA, and using the cache coherency protocol as a way to extract runtime information on the memory access patterns  of applications for automatic and dynamic tuning of the system.

\bibliographystyle{plain}
\bibliography{references}

\begin{thebibliography}{10}

\bibitem{fastmodels}
{ARM Ltd.}
\newblock {Fast Models}.
\newblock
  \url{https://developer.arm.com/tools-and-software/simulation-models/fast-models},
  August 2021.

\bibitem{sail}
Alasdair Armstrong, Thomas Bauereiss, Brian Campbell, Shaked Flur, Kathryn~E.
  Gray, Prashanth Mundkur, Robert~M. Norton, Christopher Pulte, Alastair Reid,
  Peter Sewell, Ian Stark, and Mark Wassell.
\newblock Detailed models of instruction set architectures: From pseudocode to
  formal semantics.
\newblock In {\em Proc. Automated Reasoning Workshop}, pages 23--24, April
  2018.
\newblock Two-page abstract. Proceedings available at
  \url{https://www.cl.cam.ac.uk/events/arw2018/arw2018-proc.pdf}.

\bibitem{arnold2020single}
Lukas~On Arnold and Muhsen Owaida.
\newblock Single-pass covariance matrix calculation on a hybrid fpga/cpu
  platform.
\newblock In {\em EPJ Web of Conferences}, volume 245, page 09006. EDP
  Sciences, 2020.

\bibitem{Baumann:Barrelfish:2009}
Andrew Baumann, Paul Barham, Pierre-Evariste Dagand, Tim Harris, Rebecca
  Isaacs, Simon Peter, Timothy Roscoe, Adrian Sch\"{u}pbach, and Akhilesh
  Singhania.
\newblock {The Multikernel: A New OS Architecture for Scalable Multicore
  Systems}.
\newblock In {\em Proceedings of the ACM SIGOPS 22Nd Symposium on Operating
  Systems Principles}, SOSP '09, pages 29--44, New York, NY, USA, 2009. ACM.

\bibitem{Berg:Maintaining:2009}
Thomas~B. Berg.
\newblock Maintaining i/o data coherence in embedded multicore systems.
\newblock {\em IEEE Micro}, 29(3):10--19, 2009.

\bibitem{CONDA19}
Amirali Boroumand, Saugata Ghose, Minesh Patel, Hasan Hassan, Brandon Lucia,
  Rachata Ausavarungnirun, Kevin Hsieh, Nastaran Hajinazar, Krishna~T. Malladi,
  Hongzhong Zheng, and Onur Mutlu.
\newblock Conda: Efficient cache coherence support for near-data accelerators.
\newblock In {\em Proceedings of the 46th International Symposium on Computer
  Architecture}, page 629–642, 2019.

\bibitem{Cabrera:2019:EPPO}
Anthony~M. Cabrera and Roger~D. Chamberlain.
\newblock {Exploring Portability and Performance of OpenCL FPGA Kernels on
  Intel HARPv2}.
\newblock In {\em Proceedings of the International Workshop on OpenCL},
  IWOCL'19, New York, NY, USA, 2019. Association for Computing Machinery.

\bibitem{Calciu:2019:PBerry}
Irina Calciu, Ivan Puddu, Aasheesh Kolli, Andreas Nowatzyk, Jayneel Gandhi,
  Onur Mutlu, and Pratap Subrahmanyam.
\newblock {Project PBerry: FPGA Acceleration for Remote Memory}.
\newblock In {\em Proceedings of the Workshop on Hot Topics in Operating
  Systems}, HotOS '19, page 127–135, New York, NY, USA, 2019. Association for
  Computing Machinery.

\bibitem{ccix}
{CCIX Consortium and others}.
\newblock {Cache Coherent Interconnect for Accelerators (CCIX)}.
\newblock \url{http://www.ccixconsortium.com}, January 2019.

\bibitem{Censier:Solution:1978}
Lucian~M. Censier and Paul Feautrier.
\newblock A new solution to coherence problems in multicache systems.
\newblock {\em IEEE Transactions on Computers}, C-27(12):1112--1118, 1978.

\bibitem{DeNovo11}
Byn Choi, Rakesh Komuravelli, Hyojin Sung, Robert Smolinski, Nima Honarmand,
  Sarita~V. Adve, Vikram~S. Adve, Nicholas~P. Carter, and Ching-Tsun Chou.
\newblock Denovo: Rethinking the memory hierarchy for disciplined parallelism.
\newblock In {\em 2011 International Conference on Parallel Architectures and
  Compilation Techniques}, pages 155--166, 2011.

\bibitem{Cock:Enzian:2022}
David Cock, Abishek Ramdas, Daniel Schwyn, Michael Giardino, Adam Turowski,
  Zhenhao He, Nora Hossle, Dario Korolija, Melissa Licciardello, Kristina
  Martsenko, Reto Achermann, Gustavo Alonso, and Timothy Roscoe.
\newblock Enzian: an open, general {CPU}/{FPGA} platform for systems software
  research.
\newblock In {\em Proceedings of the 27th ACM International Conference on
  Architectural Support for Programming Languages and Operating Systems},
  ASPLOS 2022, pages 590--607, New York, NY, USA, 2022. Association for
  Computing Machinery.

\bibitem{Cook:CARRV:2027}
Henry Cook, Wesley Terpstra, and Yunsup Lee.
\newblock Diplomatic design patterns: A tilelink case study.
\newblock In {\em First Workshop on Computer Architecture Research with RISC-V
  (CARRV 2017)}, 2017.

\bibitem{Intel:QPI}
Intel Corporation.
\newblock An introduction to the intel quickpath interconnect, Jan 2009.

\bibitem{cxl}
{CXL Consortium}.
\newblock {Compute Express Link}.
\newblock \url{https://www.computeexpresslink.org/}, May 2020.

\bibitem{Fang:2019:ARDA}
Yuanwei Fang, Chen Zou, and Andrew~A. Chien.
\newblock {Accelerating Raw Data Analysis with the ACCORDA Software and
  Hardware Architecture}.
\newblock {\em Proc. VLDB Endow.}, 12(11):1568–1582, July 2019.

\bibitem{gen_z}
{Gen-Z Consortium}.
\newblock {Gen-Z Core Specification 1.1}.
\newblock \url{https://genzconsortium.org/}, August 2020.

\bibitem{Giacomoni:FastForward:2008}
John Giacomoni, Tipp Moseley, and Manish Vachharajani.
\newblock Fastforward for efficient pipeline parallelism: A cache-optimized
  concurrent lock-free queue.
\newblock In {\em Proceedings of the 13th ACM SIGPLAN Symposium on Principles
  and Practice of Parallel Programming}, PPoPP '08, page 43–52, New York, NY,
  USA, 2008. Association for Computing Machinery.

\bibitem{Giri:Accelerators:2018}
Davide Giri, Paolo Mantovani, and Luca~P. Carloni.
\newblock Accelerators and coherence: An soc perspective.
\newblock {\em IEEE Micro}, 38(6):36--45, 2018.

\bibitem{GiriNOCS}
Davide Giri, Paolo Mantovani, and Luca~P. Carloni.
\newblock Noc-based support of heterogeneous cache-coherence models for
  accelerators.
\newblock In {\em 2018 Twelfth IEEE/ACM International Symposium on
  Networks-on-Chip (NOCS)}, pages 1--8, 2018.

\bibitem{Hackenberg:Comparing:2009}
Daniel Hackenberg, Daniel Molka, and Wolfgang~E. Nagel.
\newblock {Comparing Cache Architectures and Coherency Protocols on X86-64
  Multicore SMP Systems}.
\newblock In {\em Proceedings of the 42nd Annual IEEE/ACM International
  Symposium on Microarchitecture}, page 413–422, 2009.

\bibitem{Intel:Agilex}
{Intel}.
\newblock Intel agilex fpga product brief.
\newblock
  \url{https://www.intel.com/content/www/us/en/products/docs/programmable/agilex-fpga-product-brief.html},
  2020.

\bibitem{Istvan:2016:Runtime}
Zsolt Istv{\'a}n, David Sidler, and Gustavo Alonso.
\newblock Runtime parameterizable regular expression operators for databases.
\newblock In {\em 2016 IEEE 24th Annual International Symposium on
  Field-Programmable Custom Computing Machines (FCCM)}, pages 204--211. IEEE,
  2016.

\bibitem{ito2016power}
Megumi Ito and Moriyoshi Ohara.
\newblock A power-efficient fpga accelerator: Systolic array with
  cache-coherent interface for pair-hmm algorithm.
\newblock In {\em 2016 IEEE Symposium in Low-Power and High-Speed Chips (COOL
  CHIPS XIX)}, pages 1--3. IEEE, 2016.

\bibitem{Jouppi:2017:IDPA}
Norman~P. Jouppi, Cliff Young, Nishant Patil, David Patterson, Gaurav Agrawal,
  Raminder Bajwa, Sarah Bates, Suresh Bhatia, Nan Boden, Al~Borchers, Rick
  Boyle, Pierre-luc Cantin, Clifford Chao, Chris Clark, Jeremy Coriell, Mike
  Daley, Matt Dau, Jeffrey Dean, Ben Gelb, Tara~Vazir Ghaemmaghami, Rajendra
  Gottipati, William Gulland, Robert Hagmann, C.~Richard Ho, Doug Hogberg, John
  Hu, Robert Hundt, Dan Hurt, Julian Ibarz, Aaron Jaffey, Alek Jaworski,
  Alexander Kaplan, Harshit Khaitan, Daniel Killebrew, Andy Koch, Naveen Kumar,
  Steve Lacy, James Laudon, James Law, Diemthu Le, Chris Leary, Zhuyuan Liu,
  Kyle Lucke, Alan Lundin, Gordon MacKean, Adriana Maggiore, Maire Mahony,
  Kieran Miller, Rahul Nagarajan, Ravi Narayanaswami, Ray Ni, Kathy Nix, Thomas
  Norrie, Mark Omernick, Narayana Penukonda, Andy Phelps, Jonathan Ross, Matt
  Ross, Amir Salek, Emad Samadiani, Chris Severn, Gregory Sizikov, Matthew
  Snelham, Jed Souter, Dan Steinberg, Andy Swing, Mercedes Tan, Gregory
  Thorson, Bo~Tian, Horia Toma, Erick Tuttle, Vijay Vasudevan, Richard Walter,
  Walter Wang, Eric Wilcox, and Doe~Hyun Yoon.
\newblock {In-Datacenter Performance Analysis of a Tensor Processing Unit}.
\newblock In {\em Proceedings of the 44th Annual International Symposium on
  Computer Architecture}, ISCA '17, page 1–12, New York, NY, USA, 2017.
  Association for Computing Machinery.

\bibitem{Kanter:Common:2007}
David Kanter.
\newblock The common system interface: Intel’s future interconnect.
\newblock {\em Real World Technologies}, pages 1--4, 2007.

\bibitem{Katz:1985:MOSI}
R.~H. Katz, S.~J. Eggers, D.~A. Wood, C.~L. Perkins, and R.~G. Sheldon.
\newblock Implementing a cache consistency protocol.
\newblock In {\em Proceedings of the 12th Annual International Symposium on
  Computer Architecture}, ISCA '85, page 276–283, Washington, DC, USA, 1985.
  IEEE Computer Society Press.

\bibitem{Cohesion11}
John~H. Kelm, Daniel~R. Johnson, William Tuohy, Steven~S. Lumetta, and
  Sanjay~J. Patel.
\newblock Cohesion: An adaptive hybrid memory model for accelerators.
\newblock {\em {IEEE} Micro}, 31(1):42--55, 2011.

\bibitem{Kokke:TinyRegex:2021}
Kokke.
\newblock A small regex implementation in c, 2021.

\bibitem{Maas:2018:Tile-Link-GC}
Martin Maas, Krste Asanovic, and John Kubiatowicz.
\newblock A hardware accelerator for tracing garbage collection.
\newblock {\em IEEE Micro}, 39(3):38--46, 2019.

\bibitem{CoherencyScales12}
Milo M.~K. Martin, Mark~D. Hill, and Daniel~J. Sorin.
\newblock Why on-chip cache coherence is here to stay.
\newblock {\em Commun. {ACM}}, 55(7):78--89, 2012.

\bibitem{Marty:Cache:2008}
Michael~R Marty.
\newblock {\em Cache Coherence Techniques for Multicore Processors}.
\newblock PhD thesis, University of Wisconsin-Madison, 2008.

\bibitem{mughrabi2021qpr}
Abdullah~T Mughrabi, Mohannad Ibrahim, and Gregory~T Byrd.
\newblock Qpr: Quantizing pagerank with coherent shared memory accelerators.
\newblock In {\em 2021 IEEE International Parallel and Distributed Processing
  Symposium (IPDPS)}, pages 962--972. IEEE, 2021.

\bibitem{CCPrimer20}
Vijay Nagarajan, Daniel~J. Sorin, Mark~D. Hill, and David~A. Wood.
\newblock {\em A Primer on Memory Consistency and Cache Coherence: Second
  Edition}.
\newblock Morgan and Claypool, 2020.

\bibitem{Nickolls:Scalable:2008}
John Nickolls, Ian Buck, Michael Garland, and Kevin Skadron.
\newblock Scalable parallel programming with cuda: Is cuda the parallel
  programming model that application developers have been waiting for?
\newblock {\em Queue}, 6(2):40--53, 2008.

\bibitem{Harp_oliver}
Neal Oliver, Rahul~R. Sharma, Stephen Chang, Bhushan Chitlur, Elkin Garcia,
  Joseph Grecco, Aaron Grier, Nelson Ijih, Yaping Liu, Pratik Marolia, Henry
  Mitchel, Suchit Subhaschandra, Arthur Sheiman, Tim Whisonant, and Prabhat
  Gupta.
\newblock A reconfigurable computing system based on a cache-coherent fabric.
\newblock In {\em 2011 International Conference on Reconfigurable Computing and
  FPGAs}, pages 80--85, 2011.

\bibitem{Owaida:2019:DIDT}
Muhsen Owaida, Amit Kulkarni, and Gustavo Alonso.
\newblock {Distributed Inference over Decision Tree Ensembles on Clusters of
  FPGAs}.
\newblock {\em ACM Trans. Reconfigurable Technol. Syst.}, 12(4), September
  2019.

\bibitem{Owaida:2017:Centaur}
Muhsen Owaida, David Sidler, Kaan Kara, and Gustavo Alonso.
\newblock {Centaur: A framework for hybrid CPU-FPGA databases}.
\newblock In {\em 2017 IEEE 25th Annual International Symposium on
  Field-Programmable Custom Computing Machines (FCCM)}, pages 211--218. IEEE,
  2017.

\bibitem{Ranganathan:2021:WSVA}
Parthasarathy Ranganathan, Daniel Stodolsky, Jeff Calow, Jeremy Dorfman,
  Marisabel Guevara, Clinton~Wills Smullen~IV, Aki Kuusela, Raghu
  Balasubramanian, Sandeep Bhatia, Prakash Chauhan, Anna Cheung, In~Suk Chong,
  Niranjani Dasharathi, Jia Feng, Brian Fosco, Samuel Foss, Ben Gelb, Sara~J.
  Gwin, Yoshiaki Hase, Da-ke He, C.~Richard Ho, Roy~W. Huffman~Jr., Elisha
  Indupalli, Indira Jayaram, Poonacha Kongetira, Cho~Mon Kyaw, Aaron Laursen,
  Yuan Li, Fong Lou, Kyle~A. Lucke, JP~Maaninen, Ramon Macias, Maire Mahony,
  David~Alexander Munday, Srikanth Muroor, Narayana Penukonda, Eric
  Perkins-Argueta, Devin Persaud, Alex Ramirez, Ville-Mikko Rautio, Yolanda
  Ripley, Amir Salek, Sathish Sekar, Sergey~N. Sokolov, Rob Springer, Don
  Stark, Mercedes Tan, Mark~S. Wachsler, Andrew~C. Walton, David~A. Wickeraad,
  Alvin Wijaya, and Hon~Kwan Wu.
\newblock {Warehouse-Scale Video Acceleration: Co-Design and Deployment in the
  Wild}.
\newblock In {\em Proceedings of the 26th ACM International Conference on
  Architectural Support for Programming Languages and Operating Systems}, page
  600–615, New York, NY, USA, 2021. Association for Computing Machinery.

\bibitem{Ravishanicar:Cache:1983}
C~Ravishanicar and James~R Goodman.
\newblock Cache implementation for multiple microprocessors.
\newblock In {\em Proceedings of the 26th IEEE Computer Society International
  Conference (COMCON)}, 1983.

\bibitem{sadri2013energy}
Mohammadsadegh Sadri, Christian Weis, Norbert Wehn, and Luca Benini.
\newblock Energy and performance exploration of accelerator coherency port
  using xilinx zynq.
\newblock In {\em Proceedings of the 10th FPGAworld Conference}, pages 1--8,
  2013.

\bibitem{Sidler:2017:Accelerating}
David Sidler, Zsolt Istv{\'a}n, Muhsen Owaida, and Gustavo Alonso.
\newblock Accelerating pattern matching queries in hybrid cpu-fpga
  architectures.
\newblock In {\em Proceedings of the 2017 ACM International Conference on
  Management of Data}, pages 403--415, 2017.

\bibitem{Sidler:RegexRTL:2017}
David Sidler, Zsolt Istv{\'a}n, Muhsen Owaida, and Gustavo Alonso.
\newblock Open sourced regular expression matching module, 2017.

\bibitem{Stone:OpenCL:2010}
John~E Stone, David Gohara, and Guochun Shi.
\newblock Opencl: A parallel programming standard for heterogeneous computing
  systems.
\newblock {\em Computing in science \& engineering}, 12(3):66, 2010.

\bibitem{Stuecheli:2018:OpenCAPI}
J.~Stuecheli, W.~J. Starke, J.~D. Irish, L.~B. Arimilli, D.~Dreps, B.~Blaner,
  C.~Wollbrink, and B.~Allison.
\newblock {IBM POWER9 Opens up a New Era of Acceleration Enablement: OpenCAPI}.
\newblock {\em IBM J. Res. Dev.}, 62(4–5):8:1–8:8, July 2018.

\bibitem{Stuecheli:2015:CAPI}
Jeffrey Stuecheli, Bart Blaner, CR~Johns, and MS~Siegel.
\newblock {CAPI: A Coherent Accelerator Processor Interface}.
\newblock {\em IBM Journal of Research and Development}, 59(1):7:1--7:7, 2015.

\bibitem{Tang:Cache:1976}
C.~K. Tang.
\newblock Cache system design in the tightly coupled multiprocessor system.
\newblock New York, NY, USA, 1976. Association for Computing Machinery.

\bibitem{Wireshark}
{The Wireshark Team}.
\newblock Wireshark.
\newblock \url{https://www.wireshark.org}, 1998--2022.

\bibitem{van2019coherently}
Jan van Lunteren, Ronald Luijten, Dionysios Diamantopoulos, Florian
  Auernhammer, Christoph Hagleitner, Lorenzo Chelini, Stefano Corda, and
  Gagandeep Singh.
\newblock Coherently attached programmable near-memory acceleration platform
  and its application to stencil processing.
\newblock In {\em 2019 Design, Automation \& Test in Europe Conference \&
  Exhibition (DATE)}, pages 668--673. IEEE, 2019.

\bibitem{Xilinx:Zynq}
{Xilinx}.
\newblock Xilinx zynq ultrascale+ mpsoc.
\newblock
  \url{https://www.xilinx.com/products/silicon-devices/soc/zynq-ultrascale-mpsoc.html},
  2022.

\end{thebibliography}

\end{document}